\documentclass[a4paper,nofootinbib]{revtex4}
\usepackage{amssymb,amsmath,amsfonts}
\begin{document}
%--------------------------------------------------------------
\title{On the condensed matter scheme for emergent gravity and interferometry}
%--------------------------------------------------------------
\author{G. Jannes}
\affiliation{Instituto de Astrof\'{i}sica de Andaluc\'{i}a, CSIC, Camino Bajo de
Hu\'{e}tor 50, 18008 Granada, Spain}
\affiliation{Instituto de Estructura de la Materia, CSIC, Serrano 121, 28006
Madrid, Spain}
%--------------------------------------------------------------

\date{\today}

%--------------------------------------------------------------
\begin{abstract}
%--------------------------------------------------------------

An increasingly popular approach to quantum gravity rests on the idea
that gravity (and maybe electromagnetism and the other gauge fields) might be an `emergent phenomenon',
in the sense of representing a collective behaviour
resulting from a very different microscopic physics. A prominent example of this approach is the
condensed matter scheme for quantum gravity, which considers the possibility that gravity emerges as an effective low-energy phenomenon from the quantum vacuum in
a way similar to the emergence of collective excitations in condensed matter systems. This idea is supported by the observation that even in
relatively simple, weakly interacting condensed matter systems, such as
Bose-Einstein condensates, the kinematics of the low-energy excitations
(phonons) can be described by a relativistic field theory, in which the
curved background spacetime is provided by the collective behaviour of
the condensed part of the constituent atoms. In more
complicated fermionic systems, in particular $^3$He-A,
gravitational and gauge fields emerge as the low-energy bosonic degrees of
freedom together with fermionic (quasi-)matter in a similar way. All
these emergent components share surprisingly many characteristics with
their counterparts in `high-energy' physics: Einstein gravity and the
standard model of particles. This leads one to speculate that these
latter theories might really just be effective theories emerging in the
low-energy corner of a fundamental theory of quantum gravity
describing the `atoms' composing the quantum vacuum.

This condensed matter view of the quantum vacuum clearly hints that,
while the term `ether' has been discredited for about a century, quantum
gravity holds many (if not all) of the characteristics that have led people
in the past to label various hypothetical substances with the term
`ether': It is an unobserved --- and maybe even in principle unobservable
--- substance that is postulated to clear up some of the theoretical
contradictions that arise from our current description of nature, in
particular from the attempt to describe matter and gravity in a unified
way. More importantly, the condensed-matter-like background seems to provide a privileged inertial frame.

Since the last burst of enthusiasm for an ether, at the end of the 19th
century, was brought to the grave in part by the performance of a series
of important
experiments in interferometry, the suggestion then naturally arises that
maybe interferometry could also play a role in the current discussion on
quantum gravity.

In this contribution, we will highlight some aspects of this suggestion in the context of the condensed matter scheme for emergent gravity.

%--------------------------------------------------------------
\end{abstract}
%--------------------------------------------------------------
%\pacs{}
\maketitle

%--------------------------------------------------------------
\section{Introduction}
\label{S:intro}
%--------------------------------------------------------------
An intriguing theorem of mathematical physics~\cite{Visser:1997ux} shows that the
equation of motion of acoustic perturbations in a perfect fluid is described by
a d'Alembertian equation in curved spacetime: $\square \phi=0$. More precisely, consider a
non-relativistic, irrotational, inviscid and barotropic perfect fluid. Its motion
will be governed by the Euler equation, the continuity equation and an
equation of state relating density and pressure: $\rho =\rho(p)$. Then, the kinematics of linearised
perturbations
around any background solution of the equations of
motion are governed by the following d'Alembertian equation:
\begin{eqnarray}\label{alembert}%
\frac{1}{\sqrt{-g}} \partial_\mu \sqrt{-g} g^{\mu\nu} \partial_\nu \phi =0,
\end{eqnarray}%
so that these acoustic perturbations travel along the null geodesics of the effective metric $g_{\mu\nu}$. This formula is well known from relativistic field theory as the equation of
motion for a (classical or quantum) massless scalar field $\phi$ propagating in a curved spacetime. Building on this
observation, the idea developed to study certain aspects of general relativity
and quantum field theory by analogy with such perfect fluid systems (see~\cite{Barcelo:2005fc} for a recent review). To take
maximal advantage of the analogy, the  microscopic physics of the fluid system
should be well understood, theoretically and experimentally, even in regimes
where the relativistic description breaks down. Then, full calculations
based on firmly verified and controlled physics are (at least in principle)
possible, even beyond the relativistic regime. Additionally, laboratory
experiments might even become
feasible that could shed light on issues of high-energy physics. 
The paradigmatic example is that of Bose-Einstein condensates (BECs)~\cite{Dalfovo:1999zz}.
BECs fulfill all the listed conditions, and are for example considered a good
candidate for a possible future experimental detection of (phononic) Hawking
radiation~\cite{Garay:1999sk},\cite{Garay:2000jj},\cite{Barcelo:2000tg},\cite{Barcelo:2001ca}. 

When talking about analogies, it is also important to keep in mind where the
analogy breaks down. In the simple model considered up to now, there are at least two
generic limitations. The first one is that the effective or acoustic metric 
\begin{align}\label{metric}
g_{\mu \nu}=\frac{\rho}{c} 
\begin{pmatrix}
v^2-c^2 && -\vec{v}^\text{T}\\
-\vec{v} && \mathbf \openone
\end{pmatrix}
\end{align}
associated to eq.~\eqref{alembert} does not reproduce all possible general relativistic metrics, since it depends only on the background velocity vector $\mathbf v$ (which is further constrained by the requirement of irrotationality) and the speed of sound $c$ (which is related to the density $\rho$). For an analysis of how the Gross-Pitaevskii or nonlinear Schr\"{o}dinger equation describing Bose-Einstein condensates can be generalised to simulate a wider range of effective metrics, see~\cite{Barcelo:2000tg}. The
second problem regards the dynamics of the system. If one wants to extend the
analogy with general relativity beyond the kinematical aspects of quantum field
theory in a curved spacetime, there would have to be some way of emulating the
Einstein field equations. However, in a real perfect fluid such as a BEC, the inherent hydrodynamics of the system dominates and completely obfuscates any possible Einsteinian gravitational dynamics based, e.g., on Sakharov's idea of induced gravity~\cite{Sakharov:1967pk},\cite{Barcelo:2001tb},\cite{Visser:2002ew}. 

There exist ideas about how to get around this problem and study the `analogue gravitational dynamics' of a condensed matter system. For example, in~\cite{Girelli:2008gc} it was shown that the analogue gravitational dynamics of a BEC system with massive quasi-particles is (semi-)classical, i.e., can be encoded in a modified Poisson equation, while a more abstract toy model which gives rise to Nordstr\"{o}m gravity was studied in~\cite{Girelli:2008qp}. These examples show that interesting progress is being made on this front, although the models studied so far are apparently not sufficiently complex to reproduce the Einstein dynamics of general relativity. However, even if at the moment not reproducing the Einstein equations, the idea that gravity might emerge from an underlying microscopic `condensed-matter-like' quantum system has at least two additional trumps to play.

First, in more complicated fermionic systems with a Fermi-point topology, and in
particular $^3$He-A~\cite{Volovik:2003fe} (see also~\cite{Volovik:2006gt},\cite{Volovik:2008dd}), fermionic quasi-matter emerges at low energy together with effective bosonic gauge and
gravitational fields from the quantum vacuum. The construction here is slightly
more involved than in the simple case of BECs (for a rigorous derivation and detailed discussion, see~\cite{Volovik:2003fe}; \cite{Volovik:1999zs} contains an accessible introduction), but its essence can be understood as follows. 
The Fermi point is the point in momentum space where the quasi-particle energy is zero. Spatial and temporal perturbations do not destroy the Fermi point, because of its topological stability. They only lead to a general deformation of the energy spectrum near the Fermi point, determined by
\begin{eqnarray}
 g^{\mu\nu}(p_\mu-p_\mu^{(0)})(p_\nu-p_\nu^{(0)})=0~,
\end{eqnarray}
where $g^{\mu\nu}=\eta^{\lambda\sigma}e^\mu_\sigma e^\nu_\sigma$, with $e^\mu_\nu$ the tetrad or vierbein field, and $\eta^{\lambda\sigma}$ the Minkowski metric. So the dynamical change of slope in the energy spectrum near the Fermi point simulates an effective gravitational field $g^{\mu\nu}$ expressed in terms of the tetrad field $e^\mu_\nu$.
It is important to note that the effective gravitational field arises as a consequence of a perturbation of the quantum vacuum, and that this leads to a Lorentzian metric $g_{\mu\nu}$ even if the underlying system is not Lorentz invariant. The quasi-particles move along the geodesics of the effective metric $g_{\mu\nu}$. Moreover, the quasi-particles and gauge fields that emerge from such systems with Fermi-point topology show striking similarities with the ones known from the standard model of particles, including chiral or Weyl fermions and effective quantum electrodynamics: an effective electromagnetic field emerges which reflects changes in the {\it position} of the Fermi point as a consequence of a perturbation of the quantum vacuum, in a similar way 
to how the gravitational field accounts for a change in the {\it slope} of the energy spectrum near the Fermi point, see again~\cite{Volovik:2003fe}. Thus Volovik has suggested that the condensed matter
analogy might not be limited to the gravitational sector, but that by carefully
studying the topological properties of quantum vacua, this might also provide a
hint for a `theory of everything' that gives a unified description of gravity and
matter.

Second, apart from the `esthetic' issue of unifying quantum mechanics and the general theory of relativity, there is arguably at least one empirical motivation for a quantum theory of gravity: the accelerated expansion of the universe, which seems to imply some form of repulsive `dark energy' (see~\cite{Padmanabhan:2008if} for a recent review of why dark energy should be interpreted as a challenge for quantum gravity, where the argument is also made that a scenario of emergent gravity is best placed to solve this issue). The first intuition from quantum field theory to explain this mysterious repulsive force was that dark energy is simply
the energy of the quantum vacuum, which makes its entry in the Einstein field equations in the
guise of the cosmological constant~\cite{Weinberg:1988cp}. Infamously, the experimentally obtained value of the cosmological constant turned out to disagree with theoretical estimates of the 
quantum vacuum energy by more than a hundred orders of magnitude, and so this
discrepancy seems to constitute an unsurmountable barrier for such an approach. However,
if one takes the condensed matter analogy seriously, then this intuition might
prove to be right after all (see~\cite{Volovik:2004gi},\cite{Volovik:2006bh}). Indeed, the value of the quantum vacuum energy in a
condensed matter system in equilibrium is regulated by
macroscopic thermodynamic principles. The energy density $\epsilon_{\rm vac}=E_{\rm vac}/V$ of a quantum many-body system relevant for the cosmological constant problem is the `macroscopic' energy density, i.e., the total energy minus the microscopic contribution $\mu n$, with $\mu$ the chemical potential and $n$ the particle density. In other words, the vacuum energy density $\epsilon_{\rm vac}$ is obtained from the expectation value $E_{\rm vac}=<{\cal H}-\mu{\cal N}>_{\rm vac}$, with ${\cal H}$ the many-body Hamiltonian and ${\cal N}$ the number operator. The equation of state relating the energy density and the pressure of the vacuum of a quantum many-body system is then simply $\epsilon_{\rm vac}=-p_{\rm vac}$. Liquid-like systems can be in a self-sustained equilibrium without external pressure at $T=0$. So the natural value for $\epsilon_{\rm vac}$ at $T=0$ in such a system is $\epsilon_{\rm vac}=0$. At $T\neq 0$, the thermal fluctuations, or quasi-particle excitations, lead to a matter pressure $p_M$, which is compensated by a non-zero vacuum pressure such that $p_{\rm vac}+p_M=0$. The vacuum energy therefore naturally evolves towards the value $\epsilon_{\rm vac}= p_M$ in equilibrium~\cite{Barcelo:2006cs}. The microscopic constituents of the system automatically adjust to obey the macroscopic thermodynamic rule, and there is no need to know the precise microscopic constitution of the system to calculate these macroscopic quantities. Furthermore, Klinkhamer and Volovik have argued~\cite{Klinkhamer:2007pe} that, even without the assumption that the quantum vacuum is liquid-like, the experimental fact that it is highly Lorentz invariant suggests that there exists a self-tuning thermodynamic variable $q$ associated to the vacuum, which automatically readjusts itself when the vacuum is perturbed towards a new equilibrium state. This vacuum variable $q$ would therefore play a role in the quantum vacuum of the universe similar to the one of the chemical potential in a laboratory many-body system. The cosmological
constant mystery then becomes a lot less unsurmountable: From having to explain
why the cosmological constant is more than a hundred orders of magnitude smaller than its theoretically expected
value, it is reduced to having to explain why it is slightly bigger than the equilibrium value which would exactly cancel the matter contribution: $\Omega_\Lambda\approx 0.7$ versus $\Omega_M\approx 0.3$. 
 So, the condensed matter approach offers at least a
qualitative framework to understand the problem of dark energy.

To sum up, we have exposed the conceptual idea of a quantum vacuum similar
to the condensed state of a condensed matter system from which gravity emerges as
collective excitations on top of this quantum vacuum background. This
framework provides the basis for an approach to (quantum) gravity that promises
at the very least to shed an original light on some aspects of quantum gravity, such as the
problem of dark energy. However, the idea of an all-pervading condensed-matter-like `substance' from which observable physics emerges,
seems heavily reminiscent of the 19th century idea of a luminiferous ether, especially since the background quantum vacuum seems to define a privileged inertial reference frame, the `laboratory' frame in which the atoms composing the condensed matter system are (on average) at rest. This raises the question whether such an approach is not falsified a priori by the various interferometry experiments that were conducted in the recent history of physics. We will treat this question in the next section.

%--------------------------------------------------------------
\section{Interferometry at the classical level}
\label{S:interferometry_classical}
%--------------------------------------------------------------
Since the general acceptance of special relativity, the term `ether' has mainly
become associated with the few remaining crackpots who still believe that the very
foundations of relativity are wrong. Nevertheless, Einstein himself repeatedly
pointed out that special relativity never proved the ether idea to be {\it
wrong}, but only that it was {\it not needed} to explain the propagation of
light (see, e.g.,~\cite{einstein1920})\footnote{Actually, in the same lecture, Einstein also stressed that `[A]ccording to the general theory of relativity space without ether is unthinkable'.}. Another way to see this is that special relativity provides a set of rules that should be obeyed by any spacetime which obeys some basic properties (in Einstein's formulation, that of having an invariant speed in at least one inertial frame and of obeying the relativity postulate)\footnote{Note that the invariance of the speed of light in {\it all} reference frames follows from these two postulates, there is no need to include it in the first one. See also the discussion in~\cite{brown2005b}, which also discusses alternative formulations of special relativity.}, independently of the microscopic theory that underlies it (and in particular, independently of whether some sort of ether is postulated or not). In any case, one should take care not to blindly identify the
condensed matter `ether' that we discussed in the previous section with the
luminiferous ether of the 19th century. The condensed matter
ether is in a sense much more radical than the luminiferous one, since it would not only serve as the medium through which electromagnetic phenomena propagate, but it would be the substance from which electromagnetism itself, and also spacetime and (quasi-)matter, effectively arise.

Note, by the way, that these properties arguably hold for {\it all} approaches to quantum gravity, including string theory and loop quantum gravity: they postulate an unobserved -- and maybe even in principle unobservable
-- substance that is introduced to clear up some of the theoretical
contradictions that arise from our current description of nature, in
particular from the attempt to describe matter and gravity in a unified
way, even though there is strictly speaking not a single empirical fact that supports the necessity for such a theory (except possibly for the accelerated expansion of the universe or `dark energy', as we discussed in the previous section).

In any case, since condensed matter systems provide an absolute reference frame (the `laboratory frame'), one might wonder whether modern interferometry experiments of the Michelson-Morley type~\cite{michelson1887} do not falsify this kind of theory. A related question is the following. Since one of the main lessons of general relativity is diffeomorphism invariance, it is generally assumed in the relativistic community that a good quantum theory of gravity should also be diffeomorphism invariant. Diffeomorphism invariance is usually understood as encompassing a requirement for background independence (see~\cite{Smolin:2005mq} and the discussion in~\cite{Rovelli:2004tv}). In our model, we are dealing with a bi-metric system. The atoms of the condensate can be thought of as obeying the Newtonian or Galilean laws of physics of the `laboratory', since $v\ll c$ for the velocity $v$ of the condensed matter fluid. The excitations inside the condensed matter systems, on their hand, see a relativistic geometry. This relativistic geometry is in a certain sense embedded in the laboratory frame, which provides a fixed background, and hence a privileged coordinate system. Then, if background independence is apparently violated, how could such a setup lead to a diffeomorphism invariant theory? 

A naive answer to these questions is the following. An observer inside, for example, a submarine could perform an `acoustic' experiment of the Michelson-Morley type in the water surrounding him to find out whether the submarine is at rest or moving with respect to the water. An acoustic signal should here be interpreted as any signal in the form of a relativistic massless field perturbation, moving at the speed of sound $c_s$ (instead of the speed of light $c$). Then, in the simplest form of the Michelson-Morley experiment\footnote{Modern versions are of course technically much more refined than the original Michelson-Morley experiment, but the argument given here is unaffected by the concrete implementation.}, the observer would take an interferometer with two perpendicular arms of length $L$ and send acoustic signals along the arms. Mirrors at the end of the interferometer arms would reflect the signals back to a common point, where the observer looks at the interference pattern. If the interference fringes move when the interferometer is rotated, then the velocity of the submarine is not equal along the directions of both arms, and hence the observer concludes that the submarine moves with respect to the `absolute' reference frame provided by the water. Note that the interferometer arms might contract with a Lorentz factor $\gamma=(1-v^2/c^2)^{-1/2}$ with respect to the water frame, but since realistically $v\ll c_s\ll c$ (the speed of sound in water is typically around 1500 m/s), by all practical means the interferometer arms can be considered to be perfectly rigid. So the experiment should work fine, the observer can determine his state of movement with respect to the water, and diffeomorphism invariance is broken by the privileged reference frame provided by the background fluid. 

The reason why the previous argument is not valid for the case of a condensed matter background, lies in the concept of the observer. To refine the answer, we need to go back to the concept of a bi-metric theory. The exterior world is the world of the laboratory, and of its Newtonian physics, where the basic building blocks are the condensate atoms and their microscopic physics. The interior world consists of everything that can be described in terms of collective excitations (phonons, quasi-particles) emerging inside the condensed matter system. An internal observer is then somebody who is limited to the observation of the interior world, and hence to the manipulation of these collective excitations emerging inside the condensed matter system. If the observer in the submarine were really an internal observer in a condensed matter world, the building blocks available for him to build an interferometer would be made out of these same collective excitations emerging inside the condensed matter system. The fundamental signalling velocity of those building blocks would be the speed of sound $c_s$ of the fluid system instead of the speed of light $c$ of the laboratory. The Lorentz contraction undergone by the interferometer arms would then be given by an acoustic Lorentz factor $\gamma_s=(1-v^2/c_s^2)^{-1/2}$, which exactly cancels any anisotropy of the observer's movement with respect to the background, and hence leads to a null result of the interferometry experiment. To emphasise the point, let us work out this idea and rebuild the Michelson-Morley experiment from scratch for an internal observer inside a condensed matter system (see~\cite{Barcelo:2007iu} for more details). We will take the simplest case of a homogeneous medium at rest in the laboratory frame. The associated internal or effective metric is then Minkowskian. The internal observer tries to build a `quasi-interferometer' from quasi-particles and wants to obtain arms that are as rigid as possible. Will the internal observer be capable of discerning whether he is at rest or moving with respect to the condensed matter medium?

When Einstein introduced special relativity~\cite{einstein1905}, he postulated rigid bars (and perfect clocks) without further ado, although he has later repeatedly declared his dissatisfaction with the aprioriness of this assumption (see for example the discussion in~\cite{brown2005} and references therein). Phenomenologically speaking, we know that rigid structures get their rigidity from the fact that their constituents are arranged in a regular and stable way. In particular, the equilibrium distance $a_0$ between two constituents (atoms, for example) is determined by a minimum in the interaction energy $E(a)$ between both atoms:
\begin{eqnarray}
 \frac{dE}{da}=0 \; \Longrightarrow \; E_0,a_0.
\end{eqnarray}
Rigidity assumes a sharp minimum at $a_0$ in the $E(a)$ function, such that a small deviation $\delta a$ from the equilibrium distance $a_0$ would lead to a large energetic disadvantage $\Delta E$. Generally speaking, this corresponds to the requirement that the interatomic potential must be, for example, of the Lennard-Jones type. A typical example is given by the dipolar interaction between globally neutral objects composed of charged substructures. At large distances $a\gg a_0$, the forces between the objects will be approximately zero. As they are brought closer to each other, but still $a>a_0$, opposed charges will rearrange themselves and the two objects will attract each other, with a force which becomes stronger as $a$ decreases. However, if the objects are forced to a separation $a<a_0$, then the charges will redistribute and due to the repulsion between like charges, a strong repellent force between the objects will come into play. 

So assume that the internal observer disposed of charged massive `elementary quasi-particles' as building blocks which he can use to compose globally neutral constituents. Furthermore, assume that these quasi-particles interact through the use of acoustic signals moving at a velocity $c_s$, which, as we discussed in the introduction, can be described as relativistic collective excitations corresponding to a field $A_\mu$. As mentioned before, we assume that if there are different types of elementary quasi-particles, they all interact with acoustic signals propagating at the {\it same} velocity $c_s$, a condition which can be related to the principle of equivalence in general relativity.\footnote{One might argue that in the quantum gravity model described in the previous section, we of course know that all these conditions are satisfied, since the quasi-particles created by the microscopic condensed-matter-like system are what we, as internal observers, perceive as the `real' particles of the standard model, and hence their conglomerates interact through electromagnetic signals. The point is that {\it any} type of interaction with {\it any} type of charge would do, as long as there is a constant signalling speed (or at least an invariant limiting speed) and a Lennard-Jones type of potential.}

Let us then first consider a single massive quasi-particle with a charge $q$, acting as a source for the relativistic field $A_\mu$. This field will then necessarily obey the following equations (more details can be found in~\cite{Barcelo:2007iu}):
\begin{eqnarray}
& \Box A_\mu - \partial_\mu (\partial^\nu A_\nu)= j_\mu~;
\\
& j_\mu= \left\{
-q\delta^3[\vec x-\vec x(t)], q (\vec v /c_s) \delta^3[\vec x-\vec x(t)]~.
\right\}
\end{eqnarray}
For a source at rest, and taking the Lorenz gauge ($\partial^\mu A_\mu=0$), the solution is
\begin{eqnarray}
A_0 = -\frac{q}{[(x-x_0)^2+(y-y_0)^2+(z-z_0)^2]^{1/2}}~; ~~~~ A_i=0~.
\end{eqnarray}
When the source moves at a velocity $v$, for example in the $x$ direction, the solution becomes
\begin{eqnarray}
 &&A_0(x) = 
-\frac{q \gamma_s}{[(\gamma_s(x-v t) - \gamma_s x_0)^2+(y-y_0)^2+(z-z_0)^2]^{1/2}}~; \nonumber
\\
&&A_x(x)= \frac{q \gamma_s (v/c_s)}{[(\gamma_s(x-v t) - \gamma_s x_0)^2
+(y-y_0)^2+(z-z_0)^2]^{1/2}}~;
\\
&&A_y(x)=A_z(x)=0~.\nonumber
\end{eqnarray}
From this solution, we see that the fields decay faster in the $x$ direction than in the orthogonal $y$ and $z$ directions. The ratio between both decays is given by the acoustic or sonic Lorentz factor introduced earlier:
\begin{eqnarray}
\gamma_s=(1-v^2/c_s^2)^{-1/2}~.
\end{eqnarray}
Since he is building an interferometer arm, the internal observer will need to add more quasi-particles. If a single quasi-particle acquires a $\gamma_s$ factor in its direction of movement, then two particles in movement, aligned along the direction of movement, will experience a modified interaction energy potential given by $E'(a)=E(\gamma_s a)$, where the prime denotes quantities in the co-moving reference frame. As before, the equilibrium distance is given by the minimum of this potential:
\begin{eqnarray}
 0=\frac{d E'(a)}{d a} \Longrightarrow E'_0,a'_0~.
\end{eqnarray}
We obtain
\begin{eqnarray}
0=\frac{d E'(a)}{d a}= \gamma_s \frac{d E(a')}{d a'}~; ~~~~~~~~~ 
a'= \gamma_s a~,
\end{eqnarray}
which shows that the minimum now occurs when $a'=a_0$, i.e., when $a=\gamma_s^{-1} a_0$. So the `real' distance (in the sense of the distance measured in the absolute reference frame provided by the laboratory) between the two quasi-atoms has decreased by an acoustic Lorentz factor $\gamma_s$ due to their velocity with respect to the medium. Of course, $\gamma_s$ is precisely the length contraction (the contraction of an interferometer arm oriented in the direction of motion) needed for the interference pattern to remain unaffected by a uniform motion. The conclusion is then obvious: A Michelson-Morley type of experiment using a quasi-interferometer
does {\it not} allow internal observers to distinguish between rest and (uniform) movement with respect to the condensed matter medium (remarks to this effect were also made in~\cite{Volovik:2003fe} and~\cite{Liberati:2001sd}). Additionally, it could be said that a `real' Lorentz-FitzGerald contraction takes place. By `real' we here mean that it is observable in a laboratory (at least, in principle, since it is not obvious that all the conditions can be fulfilled in a real condensed matter system under experimentally observable circumstances). 

It might be interesting to note that there has always been a certain discussion about whether special relativity expresses a fundamental property of spacetime or rather in the first place a set of characteristics concerning matter and its interactions. This discussion can be traced back to the points of view of Lorentz~\cite{lorentz1895} and the separation of Einstein's original article in a kinematic and a dynamical part, has been revived by John Bell in~\cite{bell1976}, and still rages among philosophers of physics (see the recent book~\cite{brown2005b} and a strong criticism of it in~\cite{norton2008}). We would certainly not claim that our simple thought model solves this discussion, but it is clear that our model fits in nicely with the `dynamical' point of view defended most prominently by Lorentz and Bell. In our model, Lorentz invariance is perfectly valid at the effective level of the internal observer. In other words, special relativity is obtained as a natural description of the internal world as experienced by its internal inhabitants, in spite of the fact that we started out with a Newtonian or Galilean external world. So, although Lorentz invariance expresses a property of the spacetime experienced in the internal world, this seems to originate in the properties of matter, or more exactly in the way in which a spacetime description for the quasi-matter emerges from the fundamental condensate structure (see~\cite{Dreyer:2006pp} for a discussion on how this view compares to other approaches for emergent gravity, and for some speculations on how far this `internal' point of view could be taken). Extending this point of view, as already mentioned in the first paragraph of this section, one might say that special relativity describes some general aspects of {\it all} spacetimes which obey certain basic properties (in particular, the existence of an invariant signalling speed, and of course the relativity postulate), independently of the microscopic structure from which it emerges. In this sense, from the point of view of special relativity, there is no difference at all between effective or analogue spacetimes on the one hand and the `fundamental' spacetime of particle physics on the other. Special relativity does not allow to conclude whether this `fundamental' spacetime is truly fundamental in some primitive sense, or also an effective construction emerging from an underlying microscopic structure, condensed-matter-like or other.

We have explicitly worked out the case of a homogeneous condensate, in which the internal geometry is Minkowskian. But the essence of the argument is the same for more complicated situations: the internal observer has no way of connecting to the microscopic external world (at least not using only internal geometrical tools; dark energy, interpreted---as we argued before---as the quantum vacuum energy, might provide an alternative way to acquire some knowledge about the microscopic physics). Therefore, from an internal point of view, there is no way of connecting to the background structure, and hence diffeomorphism invariance is preserved. In this sense, diffeomorphism invariance might be `emergent', in the sense of being valid at the effective or internal level, but not at the fundamental level, i.e., in the external world.

In any case, it is clear that the condensed matter ether model which we introduced in the previous section, is not falsified by the null result of all known Michelson-Morley type of interferometry experiments. One way to interpret this fact is to again look back at the history of the luminiferous ether and special relativity. As Einstein himself put it, the two main lessons with respect to a possible ether theory are the following~\cite{einstein1920}. ``The special theory of relativity forbids us to assume the ether to consist of particles observable through time, but the hypothesis of ether in itself is not in conflict with the special theory of relativity. Only we must be on our guard against ascribing a state of motion to the ether.'' In the light of our condensed matter model, we could rephrase this as follows. First, the particles of which the ether is constituted are not the same as the particles of which ponderable matter is constituted. Indeed, the former are the microscopic atoms of the condensed matter system, while the latter are the quasi-particles and other collective excitations of which the internal observer disposes. Second, the internal geometry as experienced by the internal observer has no way to connect to the geometry of the external, microscopic world. 

Even if the condensed matter ether fulfills these requirements and does not contradict special relativity, there are at least two additional questions that one has to ask. The first question is whether there is any {\it need} to reinstate the ether. As we have just shown, the existence of an ether is not falsified by current experiments in interferometry, but neither is it verified. The second question is then whether there might be any way to demonstrate the existence of such an ether. We will treat this second question in the next section. With respect to the first one, the possible need to reconsider the ether, at least three reasons can be given in its support. The first reason is that, as we have argued before, all attempts at a quantum theory of gravity imply the introduction of some kind of ether. Therefore, the general arguments for the need for a quantum theory of gravity are also of application here. In particular, general relativity predicts singularities, not only in mathematically exotic cases but also in situations of physical relevance such as black holes and initial `big bang' singularities in cosmological models. It is generally assumed that a quantum theory of gravity is needed to relieve this problem and avoid the occurrence of singularities (critical assessments of the need for quantum gravity can be found, e.g., in~\cite{calender2001}, \cite{wuthrich2005}, \cite{zinkernagel2006}). The second reason can be found in the same lecture by Einstein cited above:  ``(T)here is a weighty argument to be adduced in favour of the ether hypothesis. To deny the ether is ultimately to assume that empty space has no physical qualities whatever. (...) According to the general theory of relativity space is endowed with physical qualities; in this sense, therefore, there exists an ether. According to the general theory of relativity space without ether is unthinkable; for in such space there not only would be no propagation of light, but also no possibility of existence for standards of space and time (measuring-rods and clocks), nor therefore any space-time intervals in the physical sense''~\cite{einstein1920}. A third reason to consider the reintroduction of an ether in the guise of a condensed matter system is, as we already mentioned repeatedly, that it might offer a way out of the dark energy mystery.

%--------------------------------------------------------------
\section{Quantum corrections}
\label{S:quantum}
%--------------------------------------------------------------
A typical feature of condensed matter models for emergent gravity is that Lorentz invariance is an effective description for internal observers. However, the reasoning that we set out in the previous section was based purely on geometrical arguments, i.e., valid in the low-energy limit of the theory. One might wonder if it would somehow be possible for the internal observer to experiment with sufficiently high energies, or sufficiently short wavelengths, such that the effective theory of his internal world starts to break down. Would he then see signals of the high-energy microscopic laws of the external world? We will give a few indications concerning these possibilities related to interferometry. We will separate this discussion in two parts. The first concerns the detection of possible deviations from Lorentz invariance at high energies. The second concerns the possibility of detecting quantum fluctuations of spacetime.

\subsection{Deviations from Lorentz invariance}
%--------------------------------------------------------------
In the past few years, intense experimental attention has been paid to the possibility that Lorentz invariance might be an effective low-energy phenomenon, broken at high energies~\cite{Jacobson:2005bg}. At the moment, no indication has been found that this should be the case, and actually there exist very stringent bounds on possible Lorentz violations at the Planck scale. However, we will see that care should be taken when interpreting this null result. In the traditional conjectural frameworks about quantum gravity, such as string theory and loop quantum gravity, it is not clear whether Lorentz invariance should or should not be expected to be violated at high energies. In condensed matter models, however, Lorentz invariance is a low-energy effective symmetry and hence it is definitely expected to break at high energies. Note that the traditional idea from particle physics and cosmology is that symmetry breaking takes place when the energy {\it decreases} below a certain threshold. The counterexample from the condensed matter model might be more than a curiosity, but really an indication that we need to rethink some of our traditional conceptions if we are to make further significant progress in the realm of quantum gravity. It also indicates that we should take care when postulating which elements from the currently known low-energy or `effective' physics should be fundamental at the microscopic level too.

Let us return to the two examples of condensed matter models that we discussed in the first section, and discuss the relation between Lorentz breaking and the energy scales of the system. Phenomenologically speaking, Lorentz breaking can be described simply by the following power law for the dispersion relation between the energy $E$ and the momentum $p$ (we consider massless particles and write $c$ for the invariant speed of the theory, be it a speed of light or a speed of sound):
\begin{eqnarray}
 E^2=c^2p^2+\alpha c^2p^4/p_{LV}^2 ~~(+~ \text{higher-order terms})
\end{eqnarray}
where the subscript $LV$ indicates the Lorentz violation scale, and $\alpha=\pm 1$ (we assume that uneven powers of $p$ are ruled out to lowest order, since they would lead to parity violation). %Alternatively, in units such that $\hbar=1$, we have:
%
%\begin{eqnarray}
% \omega^2=c^2k^2+\alpha c^2k^4/k_{LV}^2.
%\end{eqnarray}
%
An important remark is that there is no a priori reason to expect the Lorentz violation scale to be equal to (or even of the order of) the Planck scale.

\subsubsection{Bose-Einstein condensates}
For Bose-Einstein condensates, the dispersion relation obtained from the microscopic theory in terms of the frequency $\omega$ and the wave number $k$ is the Bogoliubov dispersion relation~\cite{Dalfovo:1999zz}
\begin{eqnarray}
 \omega^2=c^2k^2+\frac{1}{4}c^2\xi^2k^4,
\end{eqnarray}
where $\xi\equiv \hbar/mc$ is the healing length of the condensate (roughly speaking, the distance needed for the condensate to smoothen out a sharp inhomogeneity in the atomic density). The problem is that it is not obvious how to connect the Lorentz violation scale $k_{LV}=2/\xi$, or alternatively $E_{LV}=\hbar c/k_{LV}=2mc^2$, with the Planck scale of the theory. A naive reasoning could be the following. The Planck scale is the scale at which deviations from the classical picture become important. Then one might be tempted to identify the Lorentz violation scale with the Planck scale $k_P$, and so the stringent experimental bounds on Lorentz violation at the Planck scale would seemingly rule out an approach based on a BEC analogy. 

However, one should take care with this interpretation for two reasons. First, the BEC model is a model for the gravitational sector of the quantum vacuum only, and (as we already pointed out earlier) probably in the first place a toy model, so we cannot expect it to reproduce all features of the real quantum vacuum. In particular, the bosonic degrees of freedom included in the BEC model might be formed by effective coupling between fermionic degrees of freedom (through the formation of Cooper pairs, for example), or they might co-exist with other (fermionic) degrees of freedom. In both cases, information about the fermionic sector might be necessary to define and establish the hierarchy of the precise characteristic scales involved in the system. Second, even considering only the simple BEC model, already various characteristic scales can be constructed from the fundamental parameters of the microscopic theory: the Planck constant $\hbar$, the mass $m$ of the condensate atoms, their density $\rho$ (or the interatomic distance $a_0\sim \rho^{-1/3}$), and the interaction potential $U$ (for weakly interacting systems such as BECs in dilute gases, one has $U({\bf r})\approx U\delta({\bf r})$, with $U\propto a_s$, the $s$-wave scattering length; note that $a_s\ll a_0$ due to the weakness of the interaction). One can for example construct a second characteristic energy scale $E_{ch2}=\hbar c/a_0$, with $c=\sqrt{U\rho/m}$. $E_{ch2}$ can be interpreted as the energy scale at which the granularity of the vacuum becomes significant. In Bose gases, in general, $mca_0/\hbar\ll 1$, and hence $E_{LV}\ll E_{ch2}$, indicating that Lorentz violations are expected at much lower energies than the energy at which the discreteness of the vacuum becomes apparent. 

The main lesson to be drawn from this example is simply that naive dimensional estimates indicating that quantum gravity effects should be expected around `{\it the} Planck scale' $E_P=\sqrt{\hbar c^5/G}$, with $G$ the gravitational constant, are indeed naive. Different types of quantum gravity phenomenology might be characterised by different, mutually independent energy scales, which are not necessarily accessible to an internal observer who is limited to the effective low-energy physics. 

\subsubsection{Fermionic vacua}
In the Fermi-point scenario, the gravitational and gauge bosons are composites made from the fundamental fermionic degrees of freedom of the microscopic theory. Klinkhamer and Volovik~\cite{Klinkhamer:2005cp} have argued that the Planck scale should then be understood as the energy scale above which the bosonic content of the low-energy theory starts to dissolve into its fundamental fermionic components.\footnote{The argument is valid for any emergent scenario based on a fermionic vacuum where the bosons are composites made from the fermionic degrees of freedom of the vacuum, not only those with a Fermi-point topology.} Based on our experimental knowledge of the coupling constants for the various interactions, they have estimated the relation between this Planck scale $E_{Pl}$ and the Lorentz violation scale $E_{LV}$. A conservative estimate shows that at the very least $E_{Pl}/E_{LV}<10^{-8}$ (the exact proportion depends on the number of fermion families $N_F$ --- for $N_F=3$, $E_{Pl}/E_{LV}\approx 10^{-25}$ is obtained). Volovik has also argued~\cite{Volovik:2003fe} that the relation between the different characteristic scales of the theory determines the relation between the hydrodynamics and the Einstein dynamics of the theory, in the sense that hydrodynamics dominates Einstein dynamics in any system for which $E_{LV}\ll E_{Pl}$ (as is the case in $^3$He-A, and also in BECs when $E_{ch2}$ is interpreted as the Planck scale). The essence of the argument is that fermions with energies above the Lorentz violation scale contaminate the effective action for bosonic fields with non-covariant (hydrodynamic) terms. Therefore, to obtain Einstein gravity, the ultraviolet cut-off scale for the fermions must be lower than the Lorentz violation scale. So a good (condensed matter or other) model for emergent gravity would require a system in which these characteristic scales are reversed with respect to the case of $^3$He-A (and BECs): $E_{Pl}\ll E_{LV}$, in agreement with astrophysical observations and with the predictions of~\cite{Klinkhamer:2005cp}. 

In any case, the additional lesson with respect to the case of BECs might well be that, in a sufficiently complex system, the various characteristic scales could in a certain sense `conspire' to protect the effective low-energy symmetries such as Lorentz invariance, and hide the microscopic physics from a low-energy observer. Indeed, if the compositeness scale of the bosons provides a cut-off for low-energy beings such as ourselves, while the Lorentz violation scale lies at much higher energies, then the latter is suppressed from observation by at least the huge factor $E_{LV}/E_{Pl}$.

\subsection{Discreteness of spacetime}
%--------------------------------------------------------------
It has been argued by Amelino-Camelia~\cite{AmelinoCamelia:1998ax},\cite{AmelinoCamelia:1999gg}
that gravitational interferometers, once all the controllable sources of noise are compensated for, might become sufficiently sensitive to detect an irremovable source of noise: quantum fluctuations due to the discreteness or foaminess of spacetime at the Planck scale. The spectrum of this noise might allow to discriminate, at least in a generic way, between various quantum gravity scenarios. Regardless of whether this argument is realistic in technological terms, let us examine whether this idea would be applicable for the emergent scenario that we have been discussing. 

As can already be guessed from the previous discussion on high-energy deviations from Lorentz invariance, the issue depends mainly on the relation between the various characteristic scales of the theory. In the BEC model, we have seen that the Lorentz violation scale $E_{LV}$ was much smaller than the characteristic scale $E_{ch2}$ (which is related to the interatomic distance $a_0$). This implies that the granularity of the vacuum should in principle be even much harder to detect than the possible occurrence of Lorentz violations. The same argument applies to $^3$He-A. However, if Volovik's argument that a good emergent model for gravity requires $E_{Pl}\ll E_{LV}$, then it would seem that the previous implication is inverted, and that the granularity of the vacuum should be much easier to detect than a possible occurrence of Lorentz violations. This reasoning should however be taken with a grain of salt. Although in the models based on BEC and $^3$He, the Planck scale was related to the interatomic distance, and hence to the granularity of the vacuum, this need not necessarily be the case for the real quantum vacuum. If $E_{Pl}$ is the compositeness scale of the gauge bosons and $E_{LV}$ the ultraviolet cut-off scale for the fermions, as suggested in~\cite{Klinkhamer:2005cp}, then the granularity scale $E_{gran}$ could be unrelated to either of the previous two. Intuitively, one might expect $E_{gran} \geq E_{LV}$, since it seems that Lorentz invariance should certainly be violated at energies where discreteness effects become important, and maybe already at much lower energies. In that case, the granularity of the vacuum would be hidden from the low-energy physics at least as strongly as the Lorentz violation scale. However, it could also be that $E_{gran}<E_{LV}$. Indeed, there exist models which include a minimum length directly in what we would call the effective spacetime without violating Lorentz invariance~\cite{Magueijo:2001cr}. So it should not be too hard to imagine that in an emergent scenario the discreteness of the fundamental quantum vacuum could leave the Lorentz invariance of the emergent spacetime unaffected. Whether the granularity of the vacuum would then have any noticeable effect at all in the emergent spacetime remains subject for thought. An intriguing (though purely speculative) possibility is that the effective spacetime itself would not be modified, but that gravity would be modified at energies $E_{Pl}< E< E_{LV}$ and might even vanish completely before $E_{LV}$ is reached. One would then be left with a relativistic spacetime in the pure sense of special relativity~\cite{Volovik:2003nx}: a non-gravitating, but still well-defined Lorentzian spacetime, where the invariant speed $c$, which at low energy was the signalling velocity of the (massless) bosons, is now the limiting velocity of the fermions. Whether this would be detectable by means of interferometry, in a technologically feasible way or even only in principle, remains to be seen. Even so, assuming that $E_{Pl}<E_{gran}$ by at least a few orders of magnitude, this granularity would still be much harder to detect than in the conventional scenarios which handle only a single characteristic Planck scale.

%--------------------------------------------------------------
\section{Summary and conclusions}
\label{S:conclusion}
%--------------------------------------------------------------
We have briefly discussed emergent scenarios for gravity based on condensed matter models, and focused on two particular cases: Bose-Einstein condensates and $^3$He-A. As a key point in favour of this approach, we mentioned that it offers a natural interpretation for dark energy or the cosmological constant problem as the energy of the quantum vacuum. These scenarios are reminiscent of ether theories, first of all because (like all other quantum gravity conjectures) they postulate an unobserved substance which is supposed to solve the problem of unifying quantum mechanics with general relativity, but second and more importantly, because the condensed matter background on which they depend seems to define an absolute reference frame. We showed that, nevertheless, this background is not detectable by an internal observer, so that the null result of interferometry experiments such as the Michelson-Morley experiment would not be violated. We also pointed out that, in these condensed matter scenarios, Lorentz invariance is an effective, low-energy symmetry which is expected to break at high energy. However, we stressed that in scenarios for emergent gravity there is no reason to expect a single characteristic energy scale for quantum gravity (`{\it the} Planck scale'), so various aspects of quantum gravity phenomenology could be associated to different energy scales. In particular, the energy scale of Lorentz violation is expected to be many orders of magnitude higher than the cut-off scale for the effective low-energy physics, which in a theory with a fermionic vacuum could for example be the bosonic compositeness scale. Then, the effective low-energy symmetries would be protected by the proportion between these two energy scales. This would mean that any modification of the continuous effective Lorentzian spacetime at high energies would be extremely hard to detect experimentally, much harder than in the usual scenarios based on a single characteristic Planck scale for all quantum gravitational effects. This should of course not be taken as a defeatist attitude, but on the contrary as an additional stimulation to further develop the field of high-precision interferometry.

%--------------------------------------------------------------
\section*{Acknowledgements}
%--------------------------------------------------------------
I am grateful to Grisha Volovik for countless illuminating discussions, and to the Low Temperature Laboratory of the Helsinki University of Technology (TKK) for hospitality. I also thank Carlos Barcel\'{o} for joined work that served as inspiration for this work, and both him and Luis Garay for carefully reading and commenting it. This work was supported by project FIS2005-05736-C03-02 of the Spanish MEC.
%============================================================

%============================================================
%-------------------------------------------------------------
\end{document}